\documentstyle[sprocl,epsf,graphicx]{article}

\def\lessim{\mathrel{\rlap{\raise.5ex\hbox{$<$}}{\lower.5ex\hbox{$\sim$}}}}
\def\gtrsim{\mathrel{\rlap{\raise.5ex\hbox{$>$}}{\lower.5ex\hbox{$\sim$}}}}

\begin{document}


\title{COSMIC RAY SIGNATURES OF MASSIVE RELIC PARTICLES}
\author{S. SARKAR} \address{Theoretical Physics, 1 Keble Road, Oxford
OX1 3NP, UK\\ E-mail: s.sarkar@physics.ox.ac.uk} \maketitle
\abstracts{The possibility that the Fermi scale is the only
fundamental energy scale of Nature is under serious consideration at
present, yet cosmic rays may already have provided direct evidence of
new physics at a much higher scale. The recent detection of very high
energy particles with no plausible astrophysical sources suggests that
these originate from the slow decays of massive particles clustered in
the halo of our Galaxy. Such particles had in fact been predicted to
exist beforehand with mass and lifetime in the range required to
explain the observations. I discuss recent work focussing on
experimental tests of this speculative but exciting idea.}

\section{Introduction}\label{sec:Intro}

The only massive particles in the Standard Model to have survived from
the Big Bang are nucleons --- protons and (bound) neutrons --- along
with a commensurate number of electrons to yield the observed charge
neutrality of the universe.\footnote{We know now that massive relic
neutrinos contribute at least as much as the luminous component of
nucleons to the present energy density. However they are unlikely to
be the dominant component of the dark matter, based on arguments
concerning structure formation.} Considerations of primordial
nucleosynthesis restrict the nucleonic contribution to the density
parameter to $\Omega_{\rm N}\lessim0.1$ and it is widely accepted that
the dark matter in galaxies and clusters which contributes
$\Omega_{\rm DM}\gtrsim0.3$ is non-nucleonic and probably composed of
a new stable relic particle. There are many candidates for the
identity of this particle but the most popular notion is that it is
associated with the new physics beyond the Standard Model necessary to
stabilize the hierarchy between the Fermi scale,
$G_{F}^{-1/2}\simeq300$~GeV, and the Planck scale,
$G_{N}^{-1/2}\simeq10^{19}$~GeV. In particular theories of (softly
broken) low energy supersymmetry (SUSY) typically imply that the
lightest SUSY partner is a neutralino with mass of order the Fermi
scale, which is absolutely stable if the discrete symmetry termed
$R$-parity is exactly conserved. Interestingly enough the relic
abundance of such a weakly interacting particle which was in thermal
equilibrium in the early universe can account for the dark matter.

In supergravity theories, there is a new energy scale of ${\cal
O}(10^{11})$~GeV --- the geometric mean of the Fermi and Planck
scales. This is the scale of the `hidden sector' in which SUSY is
broken through gaugino condensation induced by a new strong
interaction, and communicated to the visible sector through
gravitational interactions. Following the emergence of superstrings
(for which $N=1$ supergravity is the effective field theory) it was
realised \cite{confine} that the hidden sector can also serve to
confine fractionally charged states which are a generic prediction
\cite{charge} of string theory. This avoids a serious conflict with
the unsuccessful experimental searches for fractional charges but
necessarily implies the existence of (integrally charged) bound states
with mass of ${\cal O}(10^{11})$~GeV. In a specific construction with
$SU(5)\otimes\,U(1)$ unification, it was noted \cite{crypton} that
most such states would be short-lived but that the lightest such state
would only decay through non-renormalizable operators of dimension
$\geq8$ and thus have a lifetime exceeding the age of the
universe. This introduces a new candidate for the constituent of the
dark matter --- named ``cryptons'' --- interestingly similar to
nucleons which too are bound states of fractional charges and can only
decay through non-renormalizable operators.

However, just as with nucleons, their cosmological origin is a
puzzle. If such particles were ever in thermal equilibrium their relic
abundance would have been excessive since their self-annihilations are
rather inefficient. For nucleons the problem is just the opposite and
their very existence today requires an out-of-equilibrium origin. If
the same were true of cryptons, their relic abundance may well have a
cosmologically interesting value.\footnote{It has recently been noted
\cite{wimpzilla} that particles with mass of ${\cal O}(H_{\rm
inf})\sim10^{13}$~GeV --- also dubbed ``wimpzillas'' --- can be
created with a cosmologically interesting abundance through quantum
vacuum fluctuations during inflation or during the subsequent
(re)heating process.} It is then interesting to ask what the
observational signatures of such particles might be.

Reviving an old suggestion \cite{fg80}, we recognised \cite{cryptondm}
that the most sensitive probe would be in extremely high energy cosmic
rays (EHECR), specifically in the flux of high energy neutrinos which
would necessarily be created by crypton decays. The best constraint we
obtained followed from the upper limit on deeply penetrating air
showers set by the Fly's Eye atmospheric fluorescence experiment; this
implied that such particles must have a lifetime exceeding
$\sim10^{18}$~yr if they are an important constituent of the dark
matter. As this was close to the theoretically expected lifetime in
the ``flipped'' $SU(5)$ model, I was optimistic enough to suggest in a
conference talk \cite{taup91} that `` \ldots some improvement of these
experimental sensitivities can rule out (or detect!) such particles''.

Just a few months later the Fly's Eye array detected \cite{flyseye} an
event, consistent with a proton primary, but with an energy of
$(3.0\pm0.9)\times10^{11}$~GeV. This was well above the
Greisen-Zatsepin-Kuzmin (GZK) cutoff \cite{gzk} energy of
$\sim5\times10^{10}$~GeV, beyond which resonant photopion production
losses on the cosmic microwave background should limit the propagation
distance of any such strongly interacting particle to less than about
a hundred Mpc. Over a dozen such events have been detected
subsequently by the Akeno airshower array (AGASA) as well as HiRes,
the successor to Fly's Eye, so the absence of the GZK cutoff
\cite{spectrum} is now well established. However contrary to the
expectation that such high energy particles, being essentially
undeflected by the weak intergalactic magnetic fields, should point
back to their sources, the observed distribution on the sky
\cite{isotropy} is consistent with isotropy. This is quite baffling
given that that only a few astrophysical sites (active galactic nuclei
or the extended lobes of radio galaxies) are capable of accelerating
such particles, even in principle, and there are none \cite{sources}
along the arrival directions within the propagation range. Hence it is
generally acknowledged \cite{astro} that there is no ``conventional''
astrophysical explanation for the observed EHECR.

\section{EHECR from decaying dark matter}

Faced with the above conundrum, some authors have resorted to
desperate measures, e.g. postulating that the intergalactic magnetic
field may be a thousand times stronger than usually believed, so
capable of isotropising particles from a nearby active
galaxy. However, following from our previous discussion, there is a
natural explanation \cite{bs98} for both the observed isotropy and
absence of the GZK cutoff if the EHECR originate from the decays of
metastable cryptons which are part of the dark matter.\footnote{This
was independently proposed by Berezinsky, Kachelrie{\ss} and Vilenkin
\cite{bkv97}, without, however, a specific particle candidate in
mind. Kuzmin and Rubakov \cite{kr98} also made a qualitative
suggestion that EHECR may originate from relic particle decays,
however they did not make the crucial observation that such particles
would be highly concentrated in our Galactic halo.} This is because
such particles will behave as cold dark matter (CDM) and hence cluster
in the halo of the Milky Way with a concentration $\sim10^{4}$ times
higher than the cosmic average. The local flux of EHECR will thus be
dominated by decays of cryptons in the halo, implying two distinct
observational tests of the hypothesis. First, the energy spectrum and
cosmposition (nucleons, gammas, neutrinos) beyond the GZK cutoff will
be determined \cite{bs98} essentially by the physics of crypton
decays. Second, there will be a small anisotropy \cite{aniso} in the
arrival directions of EHECR since we are located $\sim8$~kpc away from
the centre of the Galaxy and should therefore observe more particles
arriving from that direction than from the anticentre. There may also
be measurable correlations between arrival times of high energy
nucleons, gammas and neutrinos.

\subsection{Particle candidates}

As noted above, the possibility of metastable relic particles with
mass of ${\cal O}(10^{11})$~GeV had been proposed
\cite{crypton,cryptondm} {\em before} the observations of EHECR beyond
the GZK cutoff. An updated discussion \cite{ben99} of such particles
in string/M-theory confirms that cryptons are indeed favoured over
other possibilities such as the Kaluza-Klein states associated with
new compact dimensions (which are too short-lived). The most likely
candidate is still a neutral pion-like `tetron' composed of four
constituents, with a minimum lifetime of
\begin{equation}
 \tau_X \simeq {1 \over m_X} \left(\frac{M}{m_X}\right)^{10},
\label{taucrypton}
\end{equation}
where $m_X\sim10^{12-13}$~GeV, and the scale $M$ of suppression of
non-renormalizable terms is of ${\cal O}(10^{18})$~GeV.\footnote{Other
authors \cite{stringrelic} have also considered string candidates for
superheavy dark matter, and discussed \cite{hs} the confinement of
fractionally charged particles into baryon-like states in the hidden
sector and the discrete symmetries required to ensure their
longevity.} {\em Thus both the mass and lifetime of the candidate
particle are motivated by topical physical considerations.}  This is
in contrast to other proposals \cite{bkv97,kr98} where the mass scale
is not given any physical motivation and the decays are presumed to be
mediated by unspecified instanton or quantum gravity effects so as to
yield a suitably long lifetime.

\subsection{Calculation of decay spectrum}

Nevertheless all such proposals have a common phenomenology in that
regardless of the decay mechanism, the spectra of the decay products
is essentially determined by the physics of QCD fragmentation
\cite{frag} and has no major astrophysical uncertainties. In
particular given that the propagation distance in the halo is
$\lessim100$~kpc, much shorter than the GZK range of $\sim100$~Mpc,
the EHECR spectrum at Earth will be the same as the decay spectrum
(apart from the decay photons which will be degraded through
scattering on background photon fields). Of course the decay mode
(e.g. 2-body vs many-body) may well play an important role.  However
in our picture the decaying particle is a singlet under Standard Model
interactions and has a mass which significantly exceeds the Fermi
scale so the inclusive spectra of final state nucleons, photons and
neutrinos should be relatively insensitive to the precise decay
channel.

\begin{figure}[htb]
\centering
\includegraphics[width=8cm]{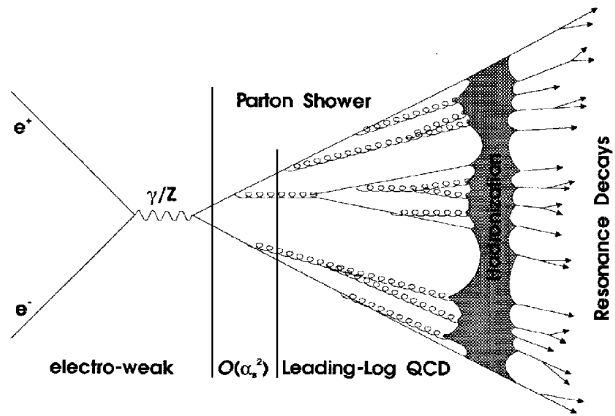}
\caption{The hadronization process \protect\cite{frag} in the decay of
 a massive particle.}
\label{fig1}
\end{figure}

We can thus imagine that we have say a $e^+e^-$ collider at our
disposal with a centre-of-mass energy $\sqrt{s}$ sufficient to create
a supermassive particle such as a crypton, rather than just a $Z^0$ as
at LEP (Figure~\ref{fig1}). This then decays into quarks and gluons
which initiate multi-parton cascades through gluon
bremsstrahlung. These finally hadronize to yield high multiplicity
jets when the momentum scale of the process drops below $\Lambda_{\rm
QCD}$. In the present context we are only interested in the final
yields of nucleons, photons and neutrinos into which all the produced
hadrons will decay. The production of different hadron species is
quantified by their respective `total fragmentation functions'
$F^h(x,s)=\sigma_{\rm tot}^{-1}{\rm d}\sigma/{\rm d}x$, viz. the
probability distributions for their inclusive production as a function
of the scaled hadron energy $x\equiv2E_h/\sqrt{s}$. These can be
factorized as the sum of contributions from different primary partons
$i=u,d,\ldots,g$:
\begin{equation}
 F^h (x, s) = \sum_i \int \frac{{\rm d}z}{z} C_i (s; z, \alpha_{\rm s}(s)) 
               D^h_i (x/z, s),
\label{factor}
\end{equation}
where $C_i$ are the `coefficient functions' dependent on the
production process, and $D_i^h$ is the `universal fragmentation
function' for parton $i\to$ hadron $j$. The essential physics in the
hard parton cascade is the logarithmic evolution of the strong
coupling $\alpha_{\rm s}$ with energy and sophisticated techniques
have been developed \cite{frag} to handle divergences associated with
collinear and soft gluon emission. The formation of the final hadrons
is however an inherently non-perturbative process and can only be
described at present by empirical models encoded in Monte Carlo event
generators, e.g. JETSET \cite{jetset} based on the `string
fragmentation' model, or HERWIG \cite{herwig} based on the `cluster
hadronization' model. These also account for the subsequent decay of
the hadrons into the observed particles, taking into account all
experimentally measured branching ratios, resonances etc, so can make
detailed predictions of measurable quantities.

\begin{figure}[htb]
\centering
\includegraphics[width=8cm]{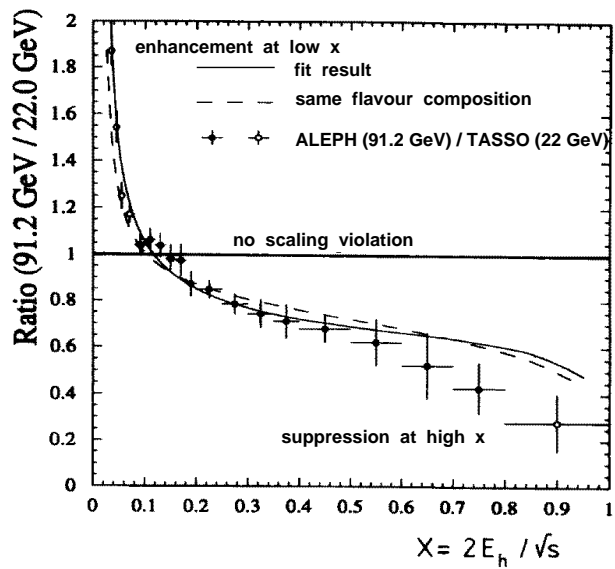}
\caption{Scale dependence of fragmentation functions
 \protect\cite{aleph} in $e^+e^-$ experiments.}
\label{scalviol}
\end{figure}

Although the fragmentation functions are not perturbatively
calculable, their evolution as a function of the momentum scale is
governed by the Dokshitzer-Gribov-Lipatov-Altarelli-Parisi (DGLAP)
equation \cite{dglap}
\begin{equation}
 s \frac{\partial}{\partial s} D_i^h (x, s) = \sum_j \int_x^1 \frac{dz}{z}
  P_{ji} (z, \alpha_{\rm s}(s)) D^h_i(x/z, s),
\label{dglap}
\end{equation}
where $P_{ji}$ are the `splitting functions' for the process parton
$i\to\,j$. Thus by measuring the fragmentation functions at one
momentum scale, one can evaluate them at another scale. As seen in
Figure~\ref{scalviol} the DGLAP equations predict violations of
`Feynmann scaling' --- a softening of the spectrum with increasing
energy --- in good agreement with data, in this case measured at PETRA
($\sqrt{s}=22$~GeV) and LEP ($\sqrt{s}=91.2$~GeV).

At small values of $x$, multiple soft gluon emission gives rise to
higher-order corrections, which turn out to be resummable by altering the
scale in the DGLAP equation (\ref{dglap}) from $s\to\,z^2s$; this 
yields a simple Gaussian function in the variable $\xi\equiv\ln(1/x)$:
\begin{equation}
 x F^h (x, s) \propto \exp 
              \left[-\frac{1}{2\sigma^2}(\xi-\xi_{\rm p})^2\right],
\label{lla}
\end{equation}
which has a characteristic peak at $\xi_{\rm p}\sim\ln\,s/4$, with
width $\sigma\sim(\ln\,s)^{3/4}$.  Including `next-to-leading'
corrections to these predictions yields the `modified leading log
approximation' (MLLA) \cite{lphd,mlla} which accounts very well for
the {\em shape} of the observed fragmentation functions at small
$x$. In comparing with data one has to further assume `local parton
hadron duality' (LPHD) \cite{lphd}, viz. that the hadron distribution
is simply proportional to the parton distribution. Thus the prediction
cannot distinguish between the individual hadronic species. Moreover
although this kinematic region dominates the total multiplicity, it
accounts for only a small fraction of the energy in the cascade, hence
the MLLA spectrum cannot be correctly normalized.

\begin{figure}[htb]
\centering
\includegraphics[width=8cm]{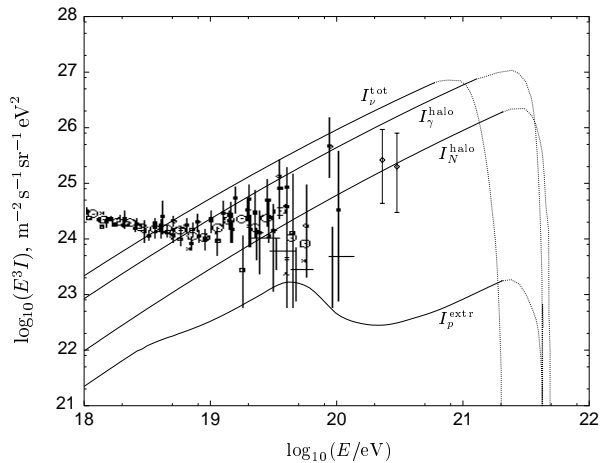}
\caption{Predicted fluxes from decaying dark matter particles of
 mass $10^{13}$~GeV according to Berezinsky {\em et al};
 \protect\cite{bkv97} the GZK-suppressed extragalactic proton flux is
 also shown.}
\label{bkvspec}
\end{figure}

With this background, we can review what has been done so far to
explain the EHECR data in terms of decaying halo particles. Berezinsky
{\em et al} \cite{bkv97} adopted the gaussian approximation
(\ref{lla}) to MLLA to infer the spectrum of nucleons from the decay
of a particle of mass $m_X$.  Although this approximation is only
valid for small $x$ ($\ll0.1$), these authors nevertheless normalized
it by requiring that $\int_0^1{\rm d}x\,xF^h(x)=f_{\rm N}$ where
$f_{\rm N}\sim0.05$ is the assumed fraction of the decaying particle
mass transferred to nucleons (on the basis of $Z^0$ decay data). The
rest is assumed to go into pions which decay to yield photons and
neutrinos, with neutral pions taking a third of the total
energy. Figure~\ref{bkvspec} shows their fit to the EHECR data
(multiplied by $E^3$ for clarity) with decaying particles of mass
$m_X=10^{13}$~GeV which contribute a fraction $\xi_X$ of the CDM
density in our halo (taken to have a radius of 100 kpc). Their adopted
normalization then implies a particle lifetime
$\tau_x/t_0=2\times10^{10}\xi_X$, where $t_0\simeq12\times10^9$~yr is
the age of the universe. Note that the constant suppression with
energy of the nucleon flux with respect to photons (and neutrinos)
follows from the {\em assumed} proportionality of the nucleon and pion
fragmentation functions.

Because of the above problems with the MLLA spectrum, we had already
considered and rejected this convenient approximation and chosen
instead to embark on a time-consuming calculation of the fragmentation
functions in the kinematic region of relevance to the data, using the
HERWIG \cite{herwig} event generator. In doing so we were initially
motivated to test an argument due to Hill \cite{h83} that the
fragmentation spectrum at large $x$ should be $\propto(1-x)^2$. By
normalizing to the total multiplicity and demanding energy
conservation Hill was then able to obtain an empirical fragmentation
function which could be fitted to extant data. Assuming the
multiplicity to be $\propto\,s^{1/4}$ as in the naive statistical
model of jet fragmentation, this was
\begin{equation}
 F_h = \frac{15}{16} x^{-3/2} (1-x)^2 ,
\label{hill1}
\end{equation}
while by adopting the leading-log QCD prediction for the multiplicity
($\propto\exp\sqrt{\ln(s/\Lambda^2)}$), he obtained
\begin{equation}
 F^h = N(b) \frac{\exp[b\sqrt{\ln(1/x)}](1-x)^2}{x\sqrt{\ln(1/x)}}.
\label{hill2}
\end{equation}
By fitting this form to PETRA data, Hill found $N(b)=0.08$ and
$b=2.6$. On the basis of the same data he assumed that 3\% of the
hadronic jets form nucleons and the other 97\% are pions which decay
into photons and neutrinos.

\begin{figure}[htb]
\centering
\includegraphics[width=8cm]{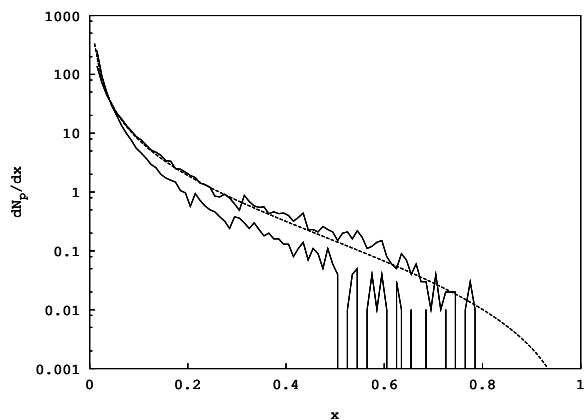}
\caption{Comparison of the nucleon fragmentation functions for
 decaying particles of mass $10^{11}$~GeV (lower curve) and
 $10^{3}$~GeV (upper curve), calculated \protect\cite{bs98} using
 HERWIG. The Hill approximation (\ref{hill2}) is also shown (dotted
 line). Note the significant scaling violation.}
\label{herwigfrag}
\end{figure}

Many authors \cite{td} who have investigated the annihilation of
GUT-scale relic topological defects (TD) as the source of EHECR have
used these expressions to estimate the fluxes. However this is clearly
inaccurate since there would be large scaling violations (see
Figure~\ref{scalviol}) in going from the PETRA energy scale of 22~GeV
up to the very much higher GUT energy scale. This is just what our
calculations \cite{bs98} using HERWIG demonstrate. The functional form
(\ref{hill2}) continues to provide a good fit to the fragmentation
function of nucleons but as shown in Figure~\ref{herwigfrag} the
spectrum becomes significantly softer with increasing particle mass,
e.g. for $m_X=10^{13}$~GeV, the normalization $N(b)$ drops to $0.0078$
(with $b=2.8$). Thus TD models \cite{td} of EHECR which use the
fragmentation functions (\ref{hill1},\ref{hill2}) {\em overestimate}
nucleon production by a factor of $\sim10$ at high energies.

\begin{figure}[hbt]
\centering
\includegraphics[width=8cm]{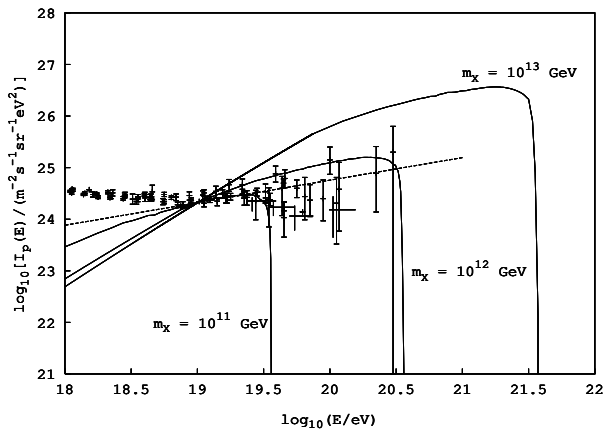}
\caption{Expected nucleon flux \protect\cite{bs98} for decaying halo
 particle masses of order the hidden sector scale compared with data
 (multiplied by $E^3$ for clarity). The spectra are normalized at
 $10^{10}$~GeV to the new component (dashed line)
 \protect\cite{flyseye} suggested by the observations.}
\label{ourspec}
\end{figure}

The EHECR spectrum in the energy range $10^{9-11}$~GeV is well fitted
\cite{flyseye} as the sum of two power-laws --- the extrapolation of
the $E^{-3.3}$ spectrum from lower energies and a new flatter
component $\propto\,E^{-2.7}$ which dominates above
$10^{10}$~GeV. (The AGASA data \cite{spectrum} gives the slope of the
new component as $-2.78^{+0.25}_{-0.33}$.)  There is some indication
\cite{flyseye} that the composition also changes from iron-group
nuclei to protons at this energy. In Figure~\ref{ourspec} we see that
the HERWIG generated spectrum \cite{bs98} for a decaying particle mass
of $10^{12}$~GeV is indeed in reasonable agreement with this new
component of cosmic rays. Our normalization requires a lifetime
$\tau_X/t_0=3\times10^{9}\xi_X$ in the notation of Berezinsky {\em et
al} \cite{bkv97}, i.e. a factor of $\sim6$ smaller. (This is because
they seem to have normalized to photons rather than nucleons at
$10^{10}$~GeV; see Figure~\ref{bkvspec}).

However our own calculations \cite{bs98} suffer from two problems. The
first, which we were initially unaware of, is that HERWIG has a known
tendency \cite{lepbar} to overproduce baryons at large $x$
(essentially due to the fragmentation of the leading quark from the
initial hard process, viz. particle decay in the present
case). Although the overall multiplicity is correctly predicted at LEP
energies (e.g. 0.953 protons per event vs $0.98\pm0.09$ observed in
$Z^0$ decay), HERWIG overproduces nucleons at $x\gtrsim0.3$ by a
factor of $\sim2-3$. Secondly, in studying the evolution of the
fragmentation function to very high energies we should take into
account that the running of the strong coupling $\alpha_{\rm s}$ would
be altered above $\sim10^3$~GeV when SUSY particles begin to be
excited from the vacuum.

\begin{figure}[htb]
\centering
\includegraphics[width=8cm]{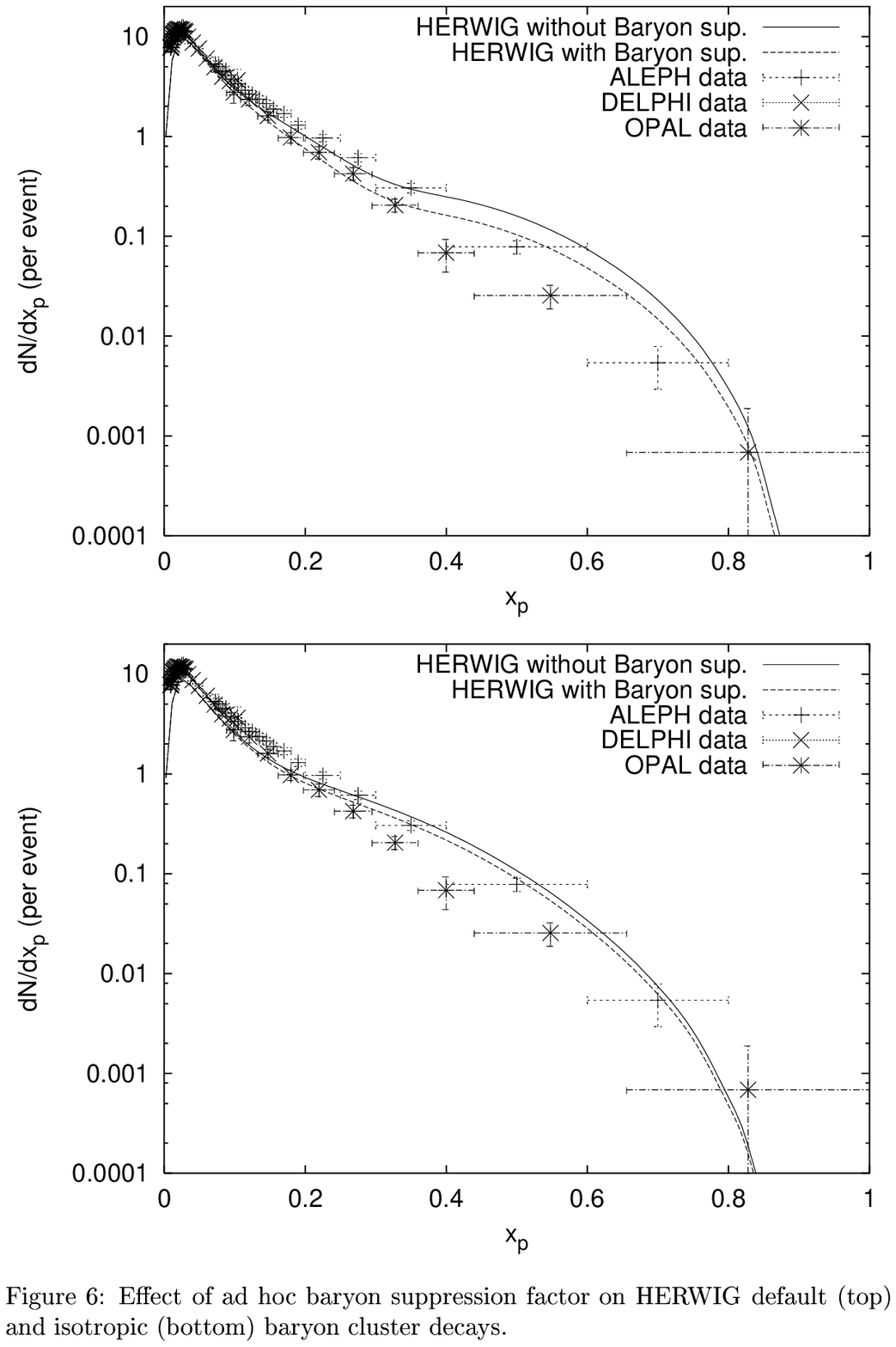}
\end{figure}

Both these issues have been addressed recently in unpublished work
\cite{r99} by Rubin. As shown in Figure~6 he finds that agreement of
HERWIG with LEP data \cite{lepbar} on baryon production is
significantly improved by isotropizing the decay of the hadron cluster
formed from the hard process (rather than have the hadron leave the
cluster along the direction of the initial quark). An additional
suppression of the probability for cluster decay by $\sim20\%$
improves the fit further. To take SUSY particles into account, he
evolves the DGLAP equation (\ref{dglap}) from LEP energies upwards
(with the initial sparton fragmentation functions calculated with the
PYTHIA \cite{jetset} event generator). The SUSY $\beta$ function for
$\alpha_{\rm s}$ is used, with the flavour thresholds corresponding to
the sparticle spectrum of a typical minimal supergravity model (with
parameters $M_0=800$~GeV, $m_{1/2}=200$~GeV, $A_0=0$, $\tan\beta=10$,
sgn$(\mu)=+$).

\begin{figure}[htb]
\centering
\includegraphics[width=8cm]{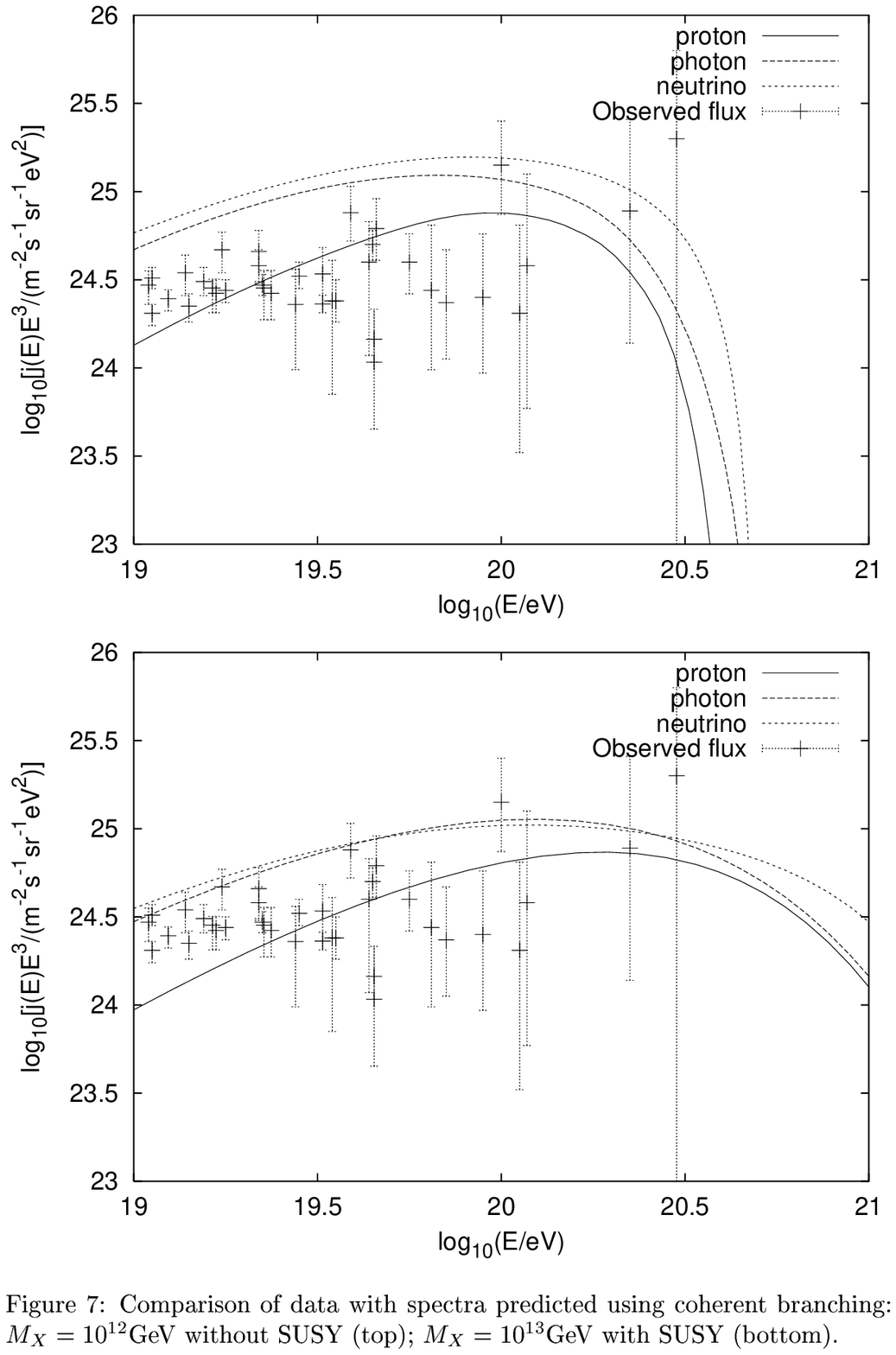}
\end{figure}

Figure~7 shows his results for both the non-SUSY and SUSY cases. The
high energy ``bump'' in our proton spectrum (see
Figure~\ref{herwigfrag}) has been erased but the non-SUSY spectrum
continues to reproduce the shape of the data for a decaying halo
particle mass of ${\cal O}(10^{12})$~GeV. The effect of including the
effects of SUSY on the evolution of the parton cascade is to flatten
the spectrum further so that a $\sim10$ times larger mass is still
acceptable. We note that the spectral shape differs considerably from
the ``SUSY-QCD'' spectrum calculated by Berezinsky and Kachelrie{\ss}
\cite{bk98} using MLLA. This is not unexpected since as emphasized
earlier, this approximation is unjustified at large $x$ so cannot be
normalized (as these authors do) to the energy released in the
decay. Moreover their assumption of an energy-independent ratio
between nucleons and pions is invalid; as is evident from Figure~7
this ratio {\em increases} with energy. \footnote{However it never
exceeds unity as in our previous work \cite{bs98} using HERWIG which
suffered from overproduction of hard baryons and gave an incorrect
prediction of the $p/\nu$ ratio.}

\section{Conclusions}

Although some progress has been made in sharpening the spectral
predictions of the decaying halo particle model for EHECR, much work
still needs to be done. The calculations so far have assumed the
simplest decay channel --- into two partons. However
non-renormalizable operators are in fact likely to induce many-body
decays.  The effects of supersymmetry also need to be investigated
more carefully, e.g. the effects of varying the SUSY parameters and
inclusion of sparticle decay channels. Nevertheless it is already
clear that the general trend in the EHECR data can be accounted for by
this hypothesis, if the particle mass is $m_X\sim10^{12-13}$~GeV and
its lifetime is $\tau_X\sim10^{16}~{\rm yr}(\xi_X/3\times10^{-4})$, so
that even with a very long lifetime such particles need constitute
only a tiny fraction $\xi_X$ of the halo CDM. It is also clear that TD
models \cite{td}, in which $m_X$ corresponds to the GUT-scale, are
already {\em ruled out} by the spectral data.

The next generation of large area cosmic ray, gamma-ray and neutrino
observatories (Auger, Amanda, Antares, \ldots) is now under
construction so it is important to refine these calculations in order
to make specific predictions for the expected fluxes. We emphasize
that previous estimates of high energy gamma-ray and neutrino fluxes
from TD \cite{tdgamnu} are based on the Hill fragmentation functions
(\ref{hill1},\ref{hill2}), while other work \cite{tdgamnu2} use the
(M)LLA spectrum (\ref{lla}) or its SUSY variant. Blasi \cite{b99} has
calculated in detail the flux of $\gamma$-rays in the decaying halo
particle model but he too uses the Hill and the MLLA spectra. All
these approximations are {\em inapplicable} at the high energies of
interest as explained earlier, and moreover the spectra of pions are
not simply proportional to that of nucleons as assumed. Hence it is
clear that all these estimates are unreliable. It is essential that
further work use the physically more realistic approach to calculating
fragmentation spectra outlined above in order to devise definitive
experimental tests \cite{us} of the decaying particle hypothesis.

\section*{Acknowledgments}
 I wish to thank the organisers of this enjoyable conference, and Neil
 Rubin for discussions and permission to quote his unpublished work.

\end{document}